\documentclass[twocolumn, showpacs, showkeys, amsmath, amssymb, superscriptaddress]{revtex4-1}

\usepackage{amsmath,amsfonts,amssymb}
\usepackage{graphicx}           

\usepackage[normalem]{ulem}     
\usepackage[usenames,dvipsnames]{color}

\begin{document}

\title{Decoherence spectroscopy with individual two-level tunneling defects}

\author{J\"urgen Lisenfeld}
    \affiliation{Physikalisches Institut, Karlsruhe Institute of Technology (KIT), 76131 Karlsruhe, Germany}
\author{Alexander Bilmes}
    \affiliation{Physikalisches Institut, Karlsruhe Institute of Technology (KIT), 76131 Karlsruhe, Germany}
\author{Shlomi Matityahu}
    \affiliation{Department of Physics, Ben-Gurion University of the Negev, Beer Sheva 84105, Israel}
\author{Sebastian Zanker}
    \affiliation{Institut f\"ur Theoretische Festk\"orperphysik, KIT, 76131 Karlsruhe, Germany}
\author{Michael Marthaler}
    \affiliation{Institut f\"ur Theoretische Festk\"orperphysik, KIT, 76131 Karlsruhe, Germany}
\author{Moshe Schechter}
    \affiliation{Department of Physics, Ben-Gurion University of the Negev, Beer Sheva 84105, Israel}\author{Gerd Sch\"on}
    \affiliation{Institut f\"ur Theoretische Festk\"orperphysik, KIT, 76131 Karlsruhe, Germany}
\author{Alexander Shnirman}
    \affiliation{Institut f\"ur Theorie der Kondensierten Materie, KIT, 76131 Karlsruhe, Germany}
    \affiliation{L. D. Landau Institute for Theoretical Physics RAS,
Kosygina street 2, 119334 Moscow, Russia}
\author{Georg Weiss}
\affiliation{Physikalisches Institut, Karlsruhe Institute of Technology (KIT), 76131 Karlsruhe, Germany}
\author{Alexey V. Ustinov}
  \affiliation{Physikalisches Institut, Karlsruhe Institute of Technology (KIT), 76131 Karlsruhe, Germany}
  \affiliation{National University of Science and Technology MISIS, Leninsky prosp. 4, Moscow, 119049, Russia}

\date{\today}

\begin{abstract}

Recent progress with microfabricated quantum devices has revealed that an ubiquitous source of noise originates in tunneling material defects that give rise to a sparse bath of parasitic two-level systems (TLSs).
For superconducting qubits, TLSs residing on electrode surfaces and in tunnel junctions account for a major part of decoherence and thus pose a serious roadblock to the realization of solid-state quantum processors.

Here, we utilize a superconducting qubit to explore the quantum
state evolution of coherently operated TLSs in order to shed new
light on their individual properties and environmental
interactions. We identify a frequency-dependence of TLS energy
relaxation rates that can be explained by a coupling to phononic
modes rather than by anticipated mutual TLS interactions.  Most investigated TLSs are found to be free of pure dephasing at their
energy degeneracy points, around which their Ramsey and spin-echo
dephasing rates scale linearly and quadratically with asymmetry
energy, respectively. We provide an explanation based on the
standard tunneling model, and identify interaction with incoherent
low-frequency (thermal) TLSs as the major mechanism of the pure
dephasing in coherent high-frequency TLS.
\end{abstract}

\maketitle
While the existence of two-level tunneling systems in amorphous
materials has been known for decades, they have attracted much
renewed interest after their detrimental effect on the performance
of microfabricated quantum devices was discovered. There is
evidence that TLSs reside in surface oxides of thin-film circuit
electrodes~\cite{Gao2008}, at disordered
interfaces~\cite{Quintana2014}, and in the tunnel barrier of
Josephson junctions~\cite{Simmonds2004}. Since TLSs possess both
electric and elastic dipole moments by which they couple to their
environment, they generate noise in various devices ranging from
microwave resonators and kinetic inductance photon
detectors~\cite{Zmuidzinas2012} through single-electron
transistors~\cite{Delsing14} to even nanomechanical
resonators~\cite{Ahn2003}. In state-of-the-art superconducting
qubits, interaction with individual TLSs constitutes a major
decoherence mechanism, where they give rise to fluctuations in
time~\cite{Clemens15} and frequency~\cite{Barends14} of qubit
relaxation rates. On the other hand, this strong interaction turns
qubits into versatile tools for studying the distribution of
TLS~\cite{Martinis05,Shalibo10}, their physical
origin~\cite{Cole10} and mutual interactions~\cite{Lisenfeld2015}
as well as their quantum dynamics~\cite{Lisenfeld2010}.

\begin{figure}[b!]
    \includegraphics[width=0.95\columnwidth]{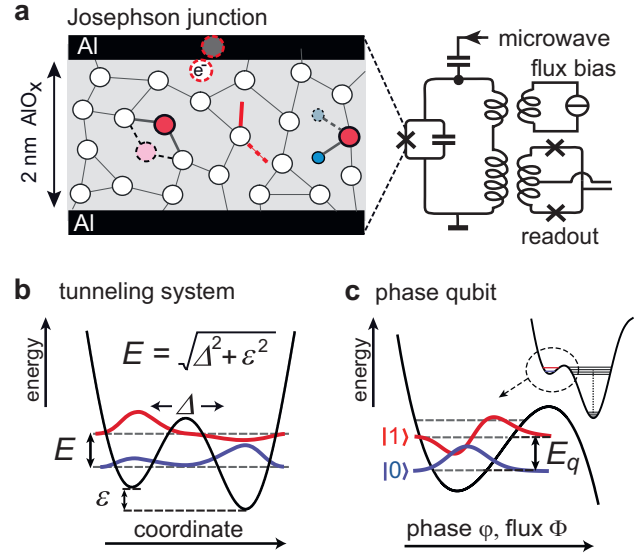}
    \caption{\textbf{Models of two-level systems (TLSs) in the Josephson junction of a superconducting qubit. a} Schematic of the phase qubit circuit used in this work and illustration of proposed TLS mechanisms: tunnelling atoms, trapped electrons, dangling bonds, and hydroxide defects.
    \textbf{b} Sketch of the TLS eigenfunctions in a double-well potential that is characterized by the strain-dependent asymmetry energy $\varepsilon$ and the tunnel coupling $\mathit{\Delta}$.
    \textbf{c} Potential energy and indication of the two lowest eigenstates of the phase qubit. }
    \label{figJL1}
\end{figure}

The omnipresence of TLSs interference is contrasted by a notable
lack of certainty regarding the microscopic nature of the
tunneling entity~\cite{Leggett2013}. Figure~\ref{figJL1}a
illustrates some proposed models of TLS formation in the amorphous
tunnel barrier of a Josephson junction: the tunnelling of
individual or small groups of atoms between two
configurations~\cite{Phillips87,PhysRevB.87.144201}, displacements of dangling bonds, and hydrogen defects~\cite{Holder2013}. Near the interface with
superconducting electrodes, TLSs may also arise from bound electron/hole Andreev states~\cite{Faoro2005} or Kondo-fluctuators~\cite{Faoro2007}.

In this work, we present first direct measurements of the decoherence rates of individual TLSs in dependence of their strain-tuned internal asymmetry energy parameter. Our experiment provides unprecedented information about the spectrum of the environment to which a TLS couples and the nature of this coupling. 

Without referring to a particular microscopic mechanism, the
standard tunnelling model~\cite{Anderson1972,Phillips1972}
assumes the potential energy of TLSs to have the form of a
double-well along a suitable configurational coordinate, giving
rise to quantum mechanical eigenstates that are superpositions of
the particle's position as illustrated in Fig.~\ref{figJL1}b).

To study the quantum state evolution of individual TLSs, we
exploit the strong interaction between a superconducting phase
qubit and defects residing in the tunnel barrier of its Josephson
junction~\cite{Simmonds2004}. Figure~\ref{figJL1}a shows the circuit
schematic of the qubit, whose potential energy is tuned via an
applied magnetic flux to adjust the energy splitting $E_q$ between
the two lowest qubit states as indicated in Fig.~\ref{figJL1}c. A TLS
is read out by tuning the unexcited qubit into resonance,
hereby realizing a coherent swap operation that maps the TLS' quantum
state onto the qubit~\cite{Lisenfeld2015}. Subsequently, a short flux pulse is applied to measure the qubit population probability
$P(|1\rangle)$~\cite{Cooper2004} which directly reflects the
population of the TLS' excited state. In our case, the TLS signal
is limited by energy relaxation that occurs in the qubit at a
characteristic time of $\Gamma_{1,\mathrm{qubit}}^{-1} \approx
100\,$ns during the readout sequence.\\

A probed TLS is characterized by the tunnelling energy $\mathit{\Delta}_p$
and the strain-dependent asymmetry energy $\varepsilon_p$ (index
$p$ stands here for ''probed''). In our experiments, we tune the
asymmetry energy \emph{in-situ} by slightly bending the sample
chip using a piezo actuator~\cite{Grabovskij2012}, resulting in
$\varepsilon_p(V_{})=\eta_p\cdot(V_{}-V_{0,p})$ where $V_{}$ is the
applied piezo voltage and $V_{0,p}$ the voltage at which the probed TLS
becomes symmetric. The coefficient $\eta_p$ is given by $\eta_p =
\gamma_p\,\partial \epsilon/\partial V_{}$, where $\gamma_p$ is the
deformation potential which indicates how strongly the probed TLS
couples to the applied strain. The strain is denoted by $\epsilon
= {\mathit{\delta} L}/{L}$ and we estimate
$\partial\epsilon/\partial V_{}\approx 10^{-6} / {\rm Volt}$ based
on results from a calibration of the piezo elongation per applied
voltage and finite-elements-simulation of the mechanical chip
deformation~\cite{Grabovskij2012}. The TLS Hamiltonian reads
\begin{eqnarray}
\hat{H}_{p} &=& \frac{1}{2} \left ( \begin{array}{cc}
\varepsilon_p(V_{}) & \Delta_p \\ \Delta_p & -\varepsilon_p(V_{})\\
\end{array} \right ) =
\frac{\varepsilon_p(V_{})}{2} \hat{\sigma}_z+ \frac{\Delta_p}{2} \hat{\sigma}_x\nonumber \\
        &=& \frac{1}{2} E_{p} \hat{\tau}_z
\label{tlshamilton}
\end{eqnarray}
with the Pauli matrices $\hat{\sigma}_x$ and $\hat{\sigma}_z$.
Diagonalization results in the energy difference between the TLS
eigenstates $E_p = \sqrt{\Delta_p^2 + \varepsilon_p(V_{})^2} =
\hbar \omega_{10}$, with Planck's constant $\hbar$ and the TLS
resonance frequency $\omega_{10}$. We defined the Pauli matrix
$\hat{\tau}_z$ in the eigenbasis of the TLS, which acts on the
eigenstates as $\hat{\tau}_z|\pm\rangle=\pm|\pm\rangle$.\\

\begin{figure}[tb!]
    \centering
    \includegraphics[width=0.9\columnwidth,height=12cm]{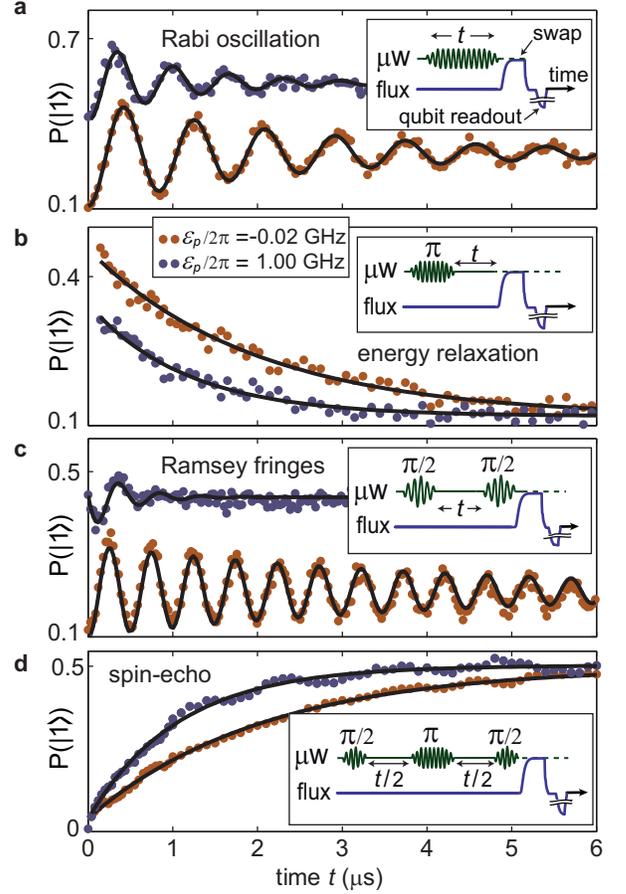}
    \caption{\textbf{Quantum dynamics of TLS3}. Each panel shows a measurement near the TLS symmetry point (red) and at $\varepsilon_p=2\pi\times 1\,$GHz (blue). Insets depict the sequence of applied microwave ($\mu$w) and flux pulses, where the latter realize a swap operation to map the TLS state onto the qubit plus a qubit readout pulse.
    \textbf{a} Rabi oscillations. \textbf{b} Energy relaxation to determine the $T_1$ time. \textbf{c} Ramsey fringes to obtain the dephasing time $T_{2,R}$. \textbf{d} Spin-echo measurement, resulting in the dephasing time $T_{2,E}$. Blue curves in \textbf{a} and \textbf{c} were shifted by 0.3 for visibility. Panels \textbf{a}-\textbf{c} show raw data of the measured qubit population probability $P(|1\rangle)$, whose reduced visibility is due to qubit energy relaxation during the TLS readout process.}
    \label{fig2}
\end{figure}

\begin{figure*}[htb!]
    \includegraphics[width=0.99\textwidth,height=11cm]{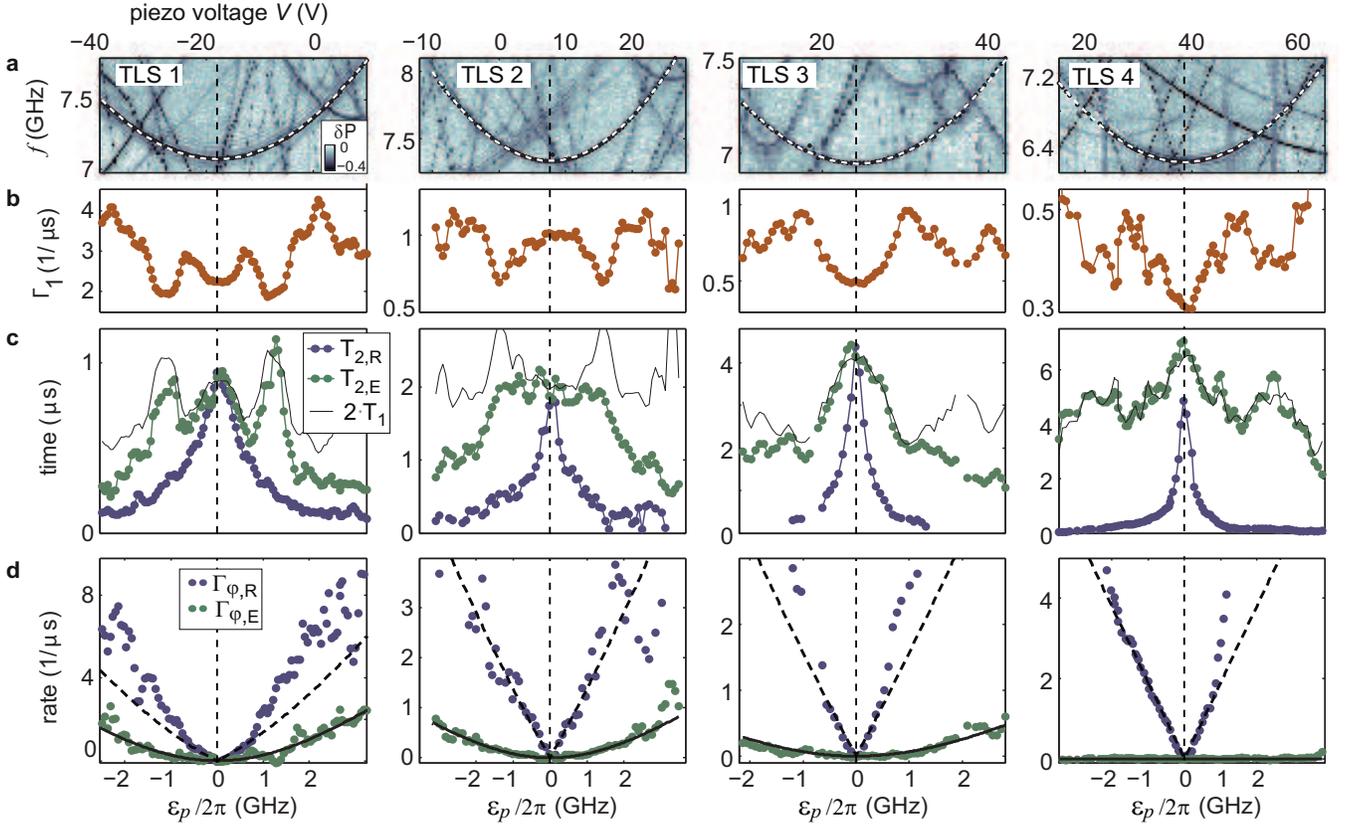}
    \caption{\textbf{Spectroscopy and results of decoherence measurements, obtained on four TLSs}. \textbf{a} Defect swap-spectroscopy, indicating the resonance frequencies of TLSs by a reduction $\delta P$ of the qubit population probability (dark traces in color-coded data). Superimposed dots are obtained from microwave spectroscopy, to which hyperbolic fits (dashed lines) result in the static TLS parameters.
    \textbf{b} Energy relaxation rate $\Gamma_1$. \textbf{c} Effective dephasing times $T_{2,R}$ (blue) and $T_{2,E}$ (green), measured using the Ramsey and spin-echo protocol, respectively. The thin black line indicates $2\cdot T_1$.
    \textbf{d} Pure dephasing rates calculated from the data in \textbf{b} and  \textbf{c}. Fitting curves (solid and dashed lines) are discussed in the text.}
    \label{figJL3}
\end{figure*}

\noindent \textbf{Experiment}\\ 
To measure TLS decoherence rates as a function of their asymmetry energy, we first apply our swap spectroscopy method~\cite{Lisenfeld2015} to obtain an overview of the TLS frequency distribution in the sample. We then select a TLS
whose symmetry point lies in the experimentally accessible strain range
and perform microwave spectroscopy to calibrate its resonance
frequency as a function of strain (see Figs.~\ref{figJL3}a). From
a hyperbolic fit, we obtain the TLS' tunnelling energy
$\mathit{\Delta_p}$, its asymmetry $\varepsilon_p(V_{})$ as a
function of the applied piezo voltage, and the deformation
potential $\gamma_p$. We then apply standard resonant microwave
pulse sequences illustrated in the insets of Fig.~\ref{fig2} to observe the
TLS' coherent state evolution in
the time domain~\cite{Lisenfeld2010}.
After a calibration of the driving strength by observing Rabi
oscillations (Fig.~\ref{fig2}a), we measure the energy relaxation
rate $\Gamma_1 \equiv 1/T_1$ by exciting the TLS with a
$\pi$-pulse and fitting the decaying state population with an
exponential $\propto \mathrm{exp}(-\Gamma_1\cdot t)$ as shown in
Fig.~\ref{fig2}b. The experimental results on $\Gamma_1$ are
summarized in Figs.~\ref{figJL3}b.

Next, we measure dephasing using the Ramsey (see Fig.~\ref{fig2}c) and the spin-echo protocol (see Fig.~\ref{fig2}d).
In both of these protocols, the TLS is initalized into a superposition of the eigenstates using a $\pi/2$-pulse.  The decay of this superposition assumes the general functional form $\exp[-\Gamma_1 t/2-x_i(t)]$, where $i=R$ for Ramsey and $i=E$ for spin-echo. The dephasing functions $x_R(t)$ and $x_E(t)$ depend strongly on the environment's fluctuation spectrum and will be in the focus of our discussion below. Further details of these experiments are contained in Supplementary Information I.

Our time-domain data do not allow us to determine the exact functional form of the dephasing signal, because only a few oscillation periods are observed at asymmetry energies where $\Gamma_{\varphi,R}$ dominates over $\Gamma_1$ (see Fig.~\ref{fig2}c). Since this renders fits to a linear dependence $x_i=\Gamma_{\varphi, i} t$ to appear practically indistinguishable from a Gaussian decay $x_i=(\Gamma_{\varphi, i} t)^2$, we estimate the pure dephasing rates $\Gamma_{\varphi, i}$ in the linear approximation and deduce their functional form from their strain dependence.
Figure set~\ref{figJL3}c summarizes the extracted effective dephasing times $T_{2,R}=(\Gamma_1/2+\Gamma_{\varphi,R})^{-1}$ and $T_{2,E}=(\Gamma_1/2+\Gamma_{\varphi,E})^{-1}$ of four different TLSs, which were measured in the same qubit sample. Using the
previously obtained data on $\Gamma_1$, we extract the pure
dephasing rates $\Gamma_{\varphi,R}$ and $\Gamma_{\varphi,E}$,
which are shown in Figs.~\ref{figJL3}d.
Additional data obtained from other TLSs are included in the Supplementary Information I.

\textbf{Energy relaxation.} For all investigated TLSs, we observe (see Figs.~\ref{figJL3}b) that their energy relaxation rates exhibit a strain-dependent structure that appears symmetric with respect to the point of lowest TLS energy $\varepsilon_p=0$.
This indicates that the spectral density of the underlying relaxing modes depends only on frequency and is independent of the applied strain.
Therefore, we conclude that the dominant relaxation mechanism of
the probed TLSs is not due to their near-resonant coupling to
other TLSs, because those would also be detuned by the applied
strain and thus are expected to generate non-symmetric patterns in $\Gamma_1$. This notion is further supported by the finding that strong mutual TLS interactions are rarely observed for our sample~\cite{Lisenfeld2015}.

If the noise spectral density was constant around $\omega =
\Delta_p / \hbar$, the strain dependence of $\Gamma_1$ for
$\varepsilon_p \ll \Delta_p$ would be given (see Supplementary Information II) by
$\Gamma_1 \propto \Delta_p^2/E_p^2 \approx 1 -
\varepsilon_p^2/(2\Delta_p^2)$, i.e.\ it would show a weak
parabolic decrease around the symmetry point. As seen in
Figs.~\ref{figJL3}b, such a scaling is obscured by the pronounced
frequency-dependence of the noise spectral density. One may assume
that this structure originates from the coupling to phonon modes which should have a discrete spectrum since the lateral size
of the junction's dielectric is comparable to the wavelength of high-frequency phonons. Indeed, a comparison of $\Gamma_1$ of different TLS as a function of their resonance frequencies (see Supplementary Information I) reveals a common maximum at 7.4 GHz for 3 out of 5 investigated TLS, indicating that those TLS may be coupled to the same phononic mode~\cite{Anghel2008}.

\textbf{Pure dephasing.} The observed pure dephasing ''rates''
$\Gamma_{\varphi,R}$ and $\Gamma_{\varphi,E}$ show the following
main features: a) the echo protocol is extraordinarily efficient, so that the ratios $\Gamma_{\varphi,R}/\Gamma_{\varphi,E}$ reach very large values (see Table~\ref{table1}). b) the $\varepsilon_p$-dependence of
the echo dephasing rate is clearly parabolic: $\Gamma_{\varphi,E}
\propto \varepsilon_p^2$; c) close to the symmetry point, the
$\varepsilon_p$-dependence of the Ramsey dephasing rate
$\Gamma_{\varphi,R}$ could be fitted to a linear behavior,
$\Gamma_{\varphi,R}\propto |\varepsilon_p|$, in all TLSs. In the Supplementary Information II, we review shortly the well known results in order to identify  possible sources of pure dephasing. We conclude that an environment characterized by white noise or by $1/f$ noise could not explain the experimental findings.

\begin{table*}
\begin{tabular}{|c|l|c|c|c|c|c|c|c|c|}   \hline \
& $\Delta_p/2\pi$  & $(\partial \varepsilon_p / \partial V) /2\pi$ & $V_{0,p}$ & $D_\parallel$ & $T_1$ @ $\varepsilon_p=0$ & $A$ & $B$ & {$\Gamma_{\varphi,R}/\Gamma_{\varphi,E}$} \\
{TLS} & (GHz) & (MHz/V) & (V) & (e\AA) & ($\mu s$) & ($\mu
s)^{-1}$ & $(\mu s)^{-1}$ & \\ \hline 
1 &  7.075 & 115.5& -18.01 & 0.37 & 0.44 & 14 & 7.7 & 8\\
2 &  7.335 & 180.3& 7.64   & 0.29 & 0.99 & 4.4 & 9.1 & 17\\
3 &  6.947 & 156.7& 24.10  & 0.26 & 2 & 3.3 & 10.5 & 22\\
4 &  6.217 & 146.8& 38.65  & 0.46 & 3.2 & 0.0 & 13.3 & $\infty$\\
\hline \end{tabular} \caption{\textbf{Measured TLS parameters.}
Static values $\Delta_p, \partial\varepsilon_p/\partial V_{}$ and
$V_{0,p}$ are obtained from a spectroscopic fit of
$\omega_{10}(V_{})$. $D_\parallel$ is the component of the TLS'
dipole moment parallel to the electric field in the junction,
extracted from the measured coupling strength to the qubit. $T_1$
is quoted at the TLS' symmetry point. Parameters $A$ and $B$ result
from fits of the measured dephasing rates in the region
$|\varepsilon_p|/2\pi<1$ GHz to the spin-echo dephasing rate
$\Gamma_{\varphi,E}=A\cdot (\varepsilon_p/E_p)^2$ and Ramsey
dephasing rate $\Gamma_{\varphi,R}=A\cdot (\varepsilon_p/E_p)^2 +
B\cdot (|\varepsilon_p|/E_p)$, respectively. The last column gives
the approximate ratio between Ramsey and echo rates, estimated in
the region $|\varepsilon_p|/2\pi<1$ GHz.}\label{table1}
\end{table*}

\textbf{Interpretation of the experimental results.} We argue that
the experimental observations can be explained in the framework of
the standard tunnelling model~\cite{Anderson1972,Phillips1972}. In
contrast to the probed high frequency TLS, whose energy splitting
is much higher than the thermal energy, $\hbar\omega_{10}\gg
k_{\rm B}T$, the TLSs responsible for pure dephasing are
''thermal'', i.e. their energy splittings are lower than $k_{\rm
B}T$ so that they switch randomly between their states. We argue
that the switching rates of thermal TLSs are very low, leading to
the essentially non-Gaussian noise that has a spectral power more
singular than $1/f$. In this case, the Ramsey dephasing is
dominated typically by the nearest neighbouring thermal
TLS~\cite{Paladino,GalperinRev04,BergliReview}. Since the
asymmetry energies of the thermal TLSs also change with strain,
one expects that some TLSs will go in and out of the group of
relevant thermal TLSs as the strain is varied. Thus, the dominant
decohering TLS will be replaced by another thermal TLS when the
change of its asymmetry energy is on the order of the thermal
energy $k_B T$ (that is, $\sim 2\pi\times 1\,$GHz). This gives
rise to a non-regular behavior, reflected in a change of slope or
small irregularities in the Ramsey dephasing rate as a function of
strain as seen in Figs.~\ref{figJL3}d. Thus, we shall focus on a
region of order $|\varepsilon_p| < k_B T \approx 2\pi\times
1\,$GHz (here and in other places we use $\hbar =1$) near the
symmetry point and study the dephasing by a single thermal TLS. In
this scenario, close to the symmetry point
$\Gamma_{\varphi,R}\propto |\varepsilon_p|$ in the general case,
or $\Gamma_{\varphi,R}\propto |\varepsilon_p|^{2}$ in the special
case where the decohering TLS is near its own symmetry point. The
dominant thermal TLS is almost completely eliminated by the echo
protocol. This explains the very high efficiency of the echo
technique. We argue that the echo dephasing rate due to the
thermal TLSs is much lower than the one due to the residual white
noise environment, which explains the observed
$\Gamma_{\varphi,E}\propto \varepsilon_p^2$.

\textbf{Theory.} In the standard tunnelling model each isolated TLS is described by a Hamiltonian $\hat{H}_{j}$ as in Eq.~(\ref{tlshamilton}), with the
index $p$ replaced by $j$. The asymmetry $\varepsilon_j$ and
tunneling energy $\Delta_j$ are assumed to be randomly distributed
with a universal distribution function $\tilde{P}(\varepsilon,\Delta)=\tilde{P}_{0}/\Delta$, where $\tilde{P}_{0}$ is a
material dependent constant~\cite{Phillips87}. Each TLS is also
characterised by its coupling $\gamma_j$ to the strain field. Moreover, it
is well established that TLSs interact via phonon-mediated
interactions, which can be described by a low-energy effective
Hamiltonian of the
form~\cite{Black:PRB:1977,esquinazi,Schechter&Stamp}
\begin{align}
\label{eq:1}&\hat{H}^{}_{\mathrm{int}}=\frac{1}{2}\,\sum^{}_{i\neq
j}J^{}_{ij}\hat{\sigma}^{}_{z,i}\hat{\sigma}^{}_{z,j}\ ,
\end{align}
with the interaction coefficients
\begin{align}
\label{eq:2}&J^{}_{ij}\sim\frac{\gamma_i \gamma_j}{\rho
c^{2}R^{3}_{ij}}\ .
\end{align}
Here $R^{}_{ij}$ is the distance between the TLSs and $\rho$, $c$
are the mass density and sound velocity, respectively. A central
dimensionless parameter of the tunnelling model is the
tunneling strength $C^{}_{0}=\tilde{P}_{0}\gamma^{2}/\rho
c^{2}=\tilde{P}_{0}\,R^{3}_{0}\,J^{}_{0}$, where $R_{0}$ and
$J_{0}=\gamma^{2}/(\rho c^{2}R^{3}_{0})$ are the typical
distance and typical interaction strength between nearest
neighbour TLSs, respectively. The well-known similarity in the
low-temperature properties of disordered solids is reflected in a universal value of $C_{0}\approx 10^{-3}$~\cite{HunklingerRev86,PohlReview}.

We consider now the probed TLS interacting with a set of thermal
TLSs via the coupling mechanism of Eq.~\eqref{eq:1}. Only the
coupling terms involving the slow (non-rotating) variables of both the probed
TLS, $\hat \tau_z$, and of the thermal TLSs, $\hat\tau_{z,j}$, are
relevant for pure dephasing. Moreover, at frequencies relevant for
pure dephasing, the operators $\hat\tau_{z,j}$ can safely be
replaced by classical stochastic processes $\tau_{z,j}(t)$
describing random switching between $\tau_{z,j} = \pm 1$ with
switching rate $\Gamma_{1,j}$. Thus, the Hamiltonian of the probed
TLS reduces effectively to
\begin{align}
\label{eq:3}&\hat{H}_{p}=\frac{1}{2}\,E_p\hat{\tau}^{}_{z}+\frac{1}{2}\,X(t)
\hat{\tau}^{}_{z}\ ,
\end{align}
where $X(t)=\sum_{j}v^{}_{j}\tau^{}_{z,j}(t)$. The effective
couplings are given by
\begin{equation}
\label{eq:vj}
v^{}_{j}=2 J^{}_{j}\cos\theta^{}_{j}\cos\theta_p \ ,
\end{equation}
where
$\cos\theta_p=\varepsilon_p/\sqrt{\varepsilon_p^2 + \Delta_p^2}$
and
$\cos\theta^{}_{j}=\varepsilon^{}_{j}/\sqrt{\varepsilon^{2}_{j}+\Delta^{2}_{j}}$.
Here $J^{}_{j}$ is the coupling strength \eqref{eq:2} between the
probed TLS and thermal TLS number $j$.

The theory of pure dephasing due to a coupling to an ensemble of
TLSs is discussed in Supplementary Information II. Here we provide the
qualitative estimates. The effect of a thermal TLS on the
coherence properties of the probed TLS depends on the coupling
$v_j$ and on the switching rate (relaxation rate) $\Gamma_{1,j}$
of the thermal TLS. We assume that the random transitions of each
TLS are mainly due to their coupling to phonons, in which case the
relaxation rate reads~\cite{Phillips1972,Jackle}
\begin{align}
\label{eq:9}&\Gamma^{}_{1,j}=
\frac{\left(2\pi\right)^{3}E_j\Delta_j^{2}\gamma_j^{2}}{\rho
h^{4}}\coth(E_j/2 k_B T)\left(\frac{1}{c^{5}_{l}}+\frac{2}{c^{5}_{t}}\right)\
,
\end{align}
where $c^{}_{l}$ and $c^{}_{t}$ are the sound velocities of the
longitudinal and transverse modes, respectively. The maximum
switching rate $\Gamma^{\text{max}}_{1,T}$ among the thermal TLS for which $E_j \le k_{\rm B}T$ (the
TLSs with $E_j>{k_{\rm B}}T$ are not thermal and do not give rise
to low frequency noise) is obtained for $\Delta_j=E_j$ and $E_j =
k_{\rm B}T$. Setting $\gamma_j\approx 1\,$eV and $T=35\,$mK~\cite{Phillips87,Shalibo10,Grabovskij2012}, we
obtain $\Gamma^{\text{max}}_{1,T}\approx
10\,$ms$^{-1}\approx 2\pi \times 1.6\, \text{kHz}$.

Next, we estimate the typical coupling strength of the nearest thermal TLS, $J_{T}$, by
calculating the typical distance between the probed TLS and its
nearest neighbouring thermal TLS in three and two dimensions (3D
and 2D, respectively), and find (see Supplementary Information II)
\begin{align}
\label{eq:11}&J_{T}=C^{}_{0}\,\xi\,k_{\rm B}T\sim 2\pi\times
10\,\text{MHz} \;\:
\left(\text{3D}\right)\,\nonumber\\
&J_{T}=C^{}_{0}\,\xi\,k_{\rm
B}T\left(\frac{d}{R^{}_{T,3D}}\right)^{3/2}\sim 2\pi\times
1\,\text{MHz} \;\: \left(\text{2D}\right)\,
\end{align}
where $\xi=\ln\left(1/u^{}_{\text{min}}\right)$, with
$u^{}_{\text{min}}$ being a lower cutoff for the parameter
$u\equiv\sin^2\theta =
\left(\Delta/E\right)^{2}$~\cite{HunklingerRev88}, $d \approx 3$ nm is the thickness of the tunnel dielectric, and we assumed
the usual values $C^{}_{0}\approx 10^{-3}$ and $\xi\approx 20$.

The above estimates reveal that thermal TLSs satisfy
$J_{T}/\Gamma^{\text{max}}_{1,T}\sim 10^3 - 10^5$. We recall the
relation $v_j =J_j \cos\theta_j \cos\theta_p$ and take into
account that, typically, $\cos\theta_j = O(1)$. Thus,
the closest thermal TLS is in the strong coupling regime, $v_j \gg
\Gamma_{1,j}$, except in the very close vicinity of the symmetry
point $\varepsilon_p=0$ of the probed TLS. Therefore, we should
study the $\varepsilon_p$-dependence of the dephasing rates
assuming the presence of strongly coupled thermal TLSs. Such a situation also provides an explanation for the effectiveness of the echo protocol. This is clearly illustrated by an example of dephasing caused by a single TLS with $v \gg \Gamma_1$ (we drop the TLS index $j$) discussed in
Ref.~\cite{GalperinRev04}. While the Ramsey dephasing ''rate'' is
of order $v$, the echo dephasing ''rate'' is of order $\Gamma_1$
of the thermal TLS (see Supplementary Information II for details). Thus, in this case,
$\Gamma_{\varphi,R}\propto |\varepsilon_p|$, whereas
$\Gamma_{\varphi,E}$ is independent of the applied strain.

The situation is more involved in the case of an ensemble of
thermal TLSs. Since the coupling strength between the TLSs scales
with their distance $r$ as $1/r^3$, the closest thermal TLS dominates the Ramsey
dephasing. Averaging the decay function over the distribution
function of TLSs is not appropriate, i.e., there
is no self-averaging (see Supplementary Information II
). The typical Ramsey decay is approximately characterized by an envelope function
$\propto \exp{\left[-\Gamma^{2}_{\varphi,R} t^2\right]}$, with possible few oscillations due to
a small number of decohering TLSs~\cite{GalperinRev04}. The Ramsey dephasing ''rate'' reads
\begin{align}
\label{eq:12}&\Gamma^{}_{\varphi,R}\approx J_{T}|\cos\theta_p|=
J_{T} \frac{|\varepsilon_p|}{E_p}\ .
\end{align}
In deriving Eq.\ \eqref{eq:12}, we assumed that the factor
$\cos\theta^{}_{j}$ in Eq.~(\ref{eq:vj}) for the closest thermal
TLS does not depend strongly on the applied strain. This
assumption is valid in the most probable case, where the closest thermal
TLS is not expected to be close to its own symmetry point
($\varepsilon_{j}=0$) at the piezo voltage $V=V^{}_{0,p}$ (i.e.\
at the same voltage for which the probed TLS is in its symmetry
point). However, in the more special case in which the closest
thermal TLS is near its symmetry point at $V=V^{}_{0,p}$, the
Ramsey dephasing rate is expected to change quadratically with the
piezo voltage, that is
\begin{align}
\label{eq:15}&\Gamma^{}_{\varphi,R}\approx
J_{T}\,\frac{\eta_p}{\Delta_p}\,\frac{\eta_j}{\Delta_j}\,\left(V-V^{}_{0}\right)^{2}\propto
\varepsilon^{2}_p\ .
\end{align}
This special situation could be of relevance for TLS1. Indeed, as one can observe in the leftmost column of Figs.~\ref{figJL3}d, a parabolic fitting could be performed here in a wider range of $\varepsilon_p$ as compared to the shown linear fit.

For the echo decay due to a single strongly coupled thermal TLS,
the dephasing rate is independent of $\varepsilon_p$. For an
ensemble of TLSs it turns out that the decay function is not
dominated by the closest TLS, but rather multiple TLSs contribute,
i.e., there is self-averaging in this case (see Supplementary Information II). 
 However, the theory predicts
$\Gamma^{}_{\varphi,E}\propto|\varepsilon_p|^{0.4}$ and
$\Gamma^{}_{\varphi,E}\propto|\varepsilon_p|^{0.5}$ in 2D and 3D,
respectively
, in disagreement with the
experimental results that show a quadratic dependence of the
echo dephasing rate. Moreover, the predicted order of magnitude is
too small to explain the experimentally observed echo dephasing
rate, i.e. the mechanism of
interactions between TLSs is expected to yield echo efficiencies
even stronger than those observed in our experiment. This is
supported by the data obtained on TLS4 (see rightmost plot of Figs.~\ref{figJL3}d), for which the echo protocol is very efficient in the whole range of
$\varepsilon_p$.

To explain the experimental findings we are forced to assume some extra
white noise environment that leads (see Supplementary Information II) to
\begin{align}
\label{eq:17}\Gamma^{}_{\varphi,E}=A\left(\frac{\varepsilon_p}{E_p}\right)^2.
\end{align}
Such a white noise environment could result e.g. from fast
relaxing TLSs~\cite{Schechter&Stamp} 
or from non-equilibrium
quasiparticles~\cite{Martinis2009,Zanker2015}. Quasiparticles are
well known to induce decoherence in superconducting quantum
devices. An estimate of the dephasing rate induced by
non-equilibrium quasiparticles is in good agreement with the
fitting parameter $A$ (see Supplementary Information II).

The above contribution of the white noise [Eq.~(\ref{eq:17})] gives
a similar contribution to the Ramsey dephasing rate. Thus,
combining (\ref{eq:12}) with (\ref{eq:17}) we attempt to fit
$\Gamma_{\varphi,R}$ in Figs.~\ref{figJL3}d in the vicinity of the
symmetry point ($|\varepsilon_p|/2\pi<1$ GHz) using
\begin{align}
\label{eq:18}\Gamma^{}_{\varphi,R}=A\left(\frac{\varepsilon_p}{E_p}\right)^2+B\,\frac{|\varepsilon_p|}{E_p}\
.
\end{align}
As explained above, one should not expect a pure linear or
parabolic behavior of the Ramsey dephasing rate on the whole range
of $\varepsilon_p$. Being dominated by single thermal TLSs, the
Ramsey dephasing rate is expected to exhibit a change of slope as
the decohering TLSs go in and out of the set of thermal TLSs, that
is when $|\varepsilon_p|/2\pi>1\,$GHz. Yet, in a typical case, one
expects a linear behavior in a narrow vicinity of the symmetry
point.

Table~\ref{table1} summarizes the fitting parameters $A$ and $B$
as well as other extracted TLS parameters. According to our
theory, the parameter $B$ is associated with the coupling $J_{T}$
[Eqs.\ \eqref{eq:11} and \eqref{eq:12}]. The estimations of the
standard tunnelling model for $J_{T}$ in the 2D case are in good agreement with the fitting parameter $B$ for all TLSs.

Summarizing, our measurements of TLS decoherence rates as a function of their asymmetry energy reveal that TLS relaxation occurs mainly due to their coupling to discrete phonon modes, while dephasing is dominated by their interaction with randomly fluctuating thermal TLS at low energies. Our theory predicts that thermal TLSs in the standard tunnelling model are characterized by $v\gg\Gamma_1$, i.e. their coupling strength to the probed TLS exceeds their switching rate. Such TLSs produce noise which gives rise to an approximately linear dependence of the Ramsey dephasing rate of coherent TLS on the external strain (which, in more special cases, can also be quadratic). The Ramsey dephasing is dominated by a small number of thermal TLSs, which explains the observed irregularities in the Ramsey dephasing rate as a function of external strain. The order of magnitude of the measured Ramsey dephasing rate is in agreement with the theory. The strain dependence of the echo dephasing rate, on the other hand, can not be accounted for by the standard tunnelling model. Its explanation requires the presence of a white noise environment. This could
consist e.g. of much faster fluctuators that are characterized by a weak
interaction with the probed TLS, or non-equilibrium quasiparticles in the
superconducting layers.

\textbf{Methods} The phase qubit sample used in this
work was fabricated in the group of J.~M. Martinis at University
of California, Santa Barbara (UCSB), as described in
Ref.~\cite{SteffenPRL2006}. The qubit junction had an area of
about $1\,\mu$m$^2$, fabricated using aluminum as electrode
material and its thermally grown oxide as a tunnel barrier. All
data have been obtained at a sample temperature of about $35\,$mK.
The mechanical strain was controlled by bending the sample chip
with a piezo transducer as explained in~\cite{Grabovskij2012}.\\

\textbf{Acknowledgements}
We would like to thank J.~M. Martinis (UCSB) for providing the
qubit sample we measured in this work. We thank A. W\"urger,
J.~H. Cole and C. M\"uller for fruitful discussions. This
work was supported by the Deutsche Forschungsgemeinschaft DFG
(Grants SCHO 287/7-1, SH 81/2-1 and LI 2446/1-1) and by the
German-Israeli Foundation (GIF Grant No. 1183-229.14/2011).
Partial support by the Ministry for Education and Science of Russian Federation under contract no. 11.G34.31.0062 and in the framework of Increase Competitiveness Program of the National University of Science and Technology MISIS under contract no. K2-2014-025 is gratefully acknowledged.
AS was supported by the Russian Science Foundation under Grant No. 14-42-00044.\\

\textbf{Author contributions} The experiments were
conceived by J.L., G.W., and A.V.U., and performed by J.L. and
A.B. Dephasing by thermal TLS within the standard tunneling model was studied by S.M., A.S., and M.S. Dephasing due to quasiparticles was analyzed by S.Z, M.M., and G.S.


\end{document}